\documentclass[sigconf]{acmart}

\copyrightyear{2026}
\acmYear{2026}
\setcopyright{cc}
\setcctype{by}
\acmConference[CHIIR '26]{2026 ACM SIGIR Conference on Human Information Interaction and Retrieval} {March 22--26, 2026}{Seattle, WA, USA}
\acmBooktitle{2026 ACM SIGIR Conference on Human Information Interaction and Retrieval (CHIIR '26), March 22--26, 2026, Seattle, WA, USA}\acmDOI{10.1145/3786304.3787866}
\acmISBN{979-8-4007-2414-5/2026/03}

\usepackage{graphicx}
\usepackage{booktabs}
\usepackage{comment}
\usepackage{subfig}
\usepackage{enumitem}
\usepackage{multirow}
\usepackage{makecell}
\usepackage{balance} 
\usepackage{pifont}

\begin{document}
\title[From Tool to Teacher: Rethinking Search Systems as Instructive Interfaces]{From Tool to Teacher: Rethinking Search Systems as Instructive Interfaces}
%\title{From Queries and Clicks to Conversations and Blends: How Users Engage with Web and Chat Systems}

    \author{David Elsweiler}
    \email{david.elsweiler@ur.de}
\affiliation{
  \institution{University of Regensburg}
  \city{Regensburg}
  \country{Germany}
  }

\renewcommand{\shortauthors}{}

\begin{abstract}
Information access systems such as search engines and generative AI are central to how people seek, evaluate, and interpret information. Yet most systems are designed to optimise retrieval rather than to help users develop better search strategies or critical awareness. This paper introduces a pedagogical perspective on information access, conceptualising search and conversational systems as instructive interfaces that can teach, guide, and scaffold users’ learning. We draw on seven didactic frameworks from education and behavioural science to analyse how existing and emerging system features, including query suggestions, source labels, and conversational or agentic AI, support or limit user learning. Using two illustrative search tasks, we demonstrate how different design choices promote skills such as critical evaluation, metacognitive reflection, and strategy transfer. The paper contributes a conceptual lens for evaluating the instructional value of information access systems and outlines design implications for technologies that foster more effective, reflective, and resilient information seekers.

\end{abstract}

\begin{comment}

\begin{CCSXML}
<ccs2012>
<concept>
<concept_id>10002951.10003317.10003331</concept_id>
<concept_desc>Information systems~Users and interactive retrieval</concept_desc>
<concept_significance>500</concept_significance>
</concept>
<concept>
<concept_id>10003120.10003121.10011748</concept_id>
<concept_desc>Human-centered computing~Empirical studies in HCI</concept_desc>
<concept_significance>500</concept_significance>
</concept>
</ccs2012>
\end{CCSXML}

\ccsdesc[500]{Information systems~Users and interactive retrieval}
\ccsdesc[500]{Human-centered computing~Empirical studies in HCI}

\keywords{search behaviour, misinformation, mixed methods}
\end{comment}


\maketitle

\section{Introduction}\label{sec:introduction}

People increasingly rely on information access systems, including  traditional search engines and generative AI (Gen-AI) interfaces, to seek knowledge, make decisions, and form beliefs, yet the ways in which users interact with these systems differ substantially. Decades of research on interactive information retrieval (IIR) shows that users vary widely in how they search and evaluate information \cite{aula2003query,white2009characterizing,rieh2000interaction,kattenbeck2019understanding, bateman2012search}. These behavioural differences are not trivial: they fundamentally shape what users find \cite{white2007investigating,zhang2005domain}, what they believe \cite{rieger2024disentangling,elsweiler2025query,mayerhofer2025blending}, and how they act \cite{west2013cookies}, with real-world consequences in areas such as health, politics, and climate.

Despite growing evidence of behavioural variability and its effects, most information access systems are still designed around a set of core assumptions: that users know what they are looking for, that they can describe it in the form of a query, prompt or conversational utterance, and that they will recognise relevant and credible information when they see it. These assumptions may hold for some well-defined tasks (e.g., "what was the acceptance rate for long papers at CHIIR last year?'' or "Where is the nearest cafe?"), but they break down in more complex, ambiguous, or high-stakes contexts, e.g., evaluating competing claims \cite{rieger2024responsible,draws2023viewpoint}, making health decisions \cite{white2013beliefs,bink2022featured}, or navigating politically charged topics \cite{lewandowsky2024truth}. In these cases, the ability to formulate effective queries, critically evaluate sources, and review one's strategy becomes just as crucial as the system's ability to retrieve relevant content. Yet, these skills are rarely supported or developed within the interaction itself.

Viewing information access systems from a pedagogical perspective allows us to conceptualise the features of the system not only in terms of usability or relevance, but also in terms of the instructional support they offer to users: in developing search strategies, evaluating source credibility, building topic understanding and reflecting on their own information behaviours. This opens up new questions: not just whether users succeed in finding information, but whether they become more effective and reflective information seekers. This perspective also encourages more intentional design of learning-oriented features, and helps reframe evaluation to include learning outcomes, critical engagement, and strategy transfer.

To advance this view, we bring together insights from behavioural science and learning theory, including nudging \cite{thaler2008nudge}, boosting \cite{hertwig2017nudging}, scaffolding and the Zone of Proximal Development (ZPD)\cite{vygotsky1978mind,wood1976role}, cognitive apprenticeship \cite{collins1989cognitive}, instructional feedback \cite{hattie2007power}, self-regulated learning \cite{zimmerman2002becoming,pintrich2000role}, and heutagogy \cite{hase2000from}. These frameworks each offer distinct but complementary ways of understanding how systems can guide, support, or empower users, either subtly, explicitly, or through situated interaction. Importantly, they also shift how we think about evaluation: not just whether a system returns relevant documents, but whether it helps users develop transferable skills, form more informed beliefs, and resist manipulation or bias.

This paper proposes that information access systems should be seen not just as retrieval engines but as didactic environments with the potential to teach, guide, and scaffold. Through this lens, we analyse common features (e.g., query suggestions, source labels) and emerging tools (e.g., Gen-AI based assistants and agentic search systems) across two illustrative search tasks. We apply a set of instructional frameworks to evaluate the latent educational roles and trade-offs in the design of these systems, and present a series of design examples that show how these frameworks can be used to build more pedagogically aware systems. We also argue for evaluation practices that account for users' long-term development, not just their immediate performance or satisfaction.

\section{Related Work}\label{sec:related_work}

Learning has not traditionally been a central concern in interactive information retrieval (IIR), yet many contributions have implicitly, or occasionally explicitly, designed or evaluated interfaces that shape user behaviour during search. We review representative examples in Section~\ref{sec:implicit_instr}. A more direct connection to learning emerges in the Search as Learning (SAL) literature, now a prominent topic within IIR. Key developments are discussed in Section~\ref{sec:sal}.

\subsection{Implicit Instruction in IIR}
\label{sec:implicit_instr}

IIR research has shown that interface design can influence search behaviour, especially in query formulation. For example, modifying the size or labelling of the search box can lead users to submit longer, more specific queries, with the goal of improving result quality \cite{belkin2003query,franzen2000verbosity}. Yamamoto and Yamamoto \cite{yamamoto2011enhancing} show that query priming via suggestions and auto-completion encourages more careful, evidence-oriented search behaviour, with effects varying by users’ educational background. Other studies have highlighted the instructional potential of feedback. Harvey et al.~\cite{harvey2015learning} demonstrated that showing users examples of effective queries, combined with immediate performance feedback %(e.g., average precision scores in a game-like setting),
helped them internalise successful patterns and improve subsequent queries. Feedback acted as a key mechanism for reflection and learning. Similarly, Bateman et al.~\cite{bateman2012search} introduced the \textit{Search Dashboard}, which visualised users’ behaviours over time and compared them to expert benchmarks. In a five-week study, participants exposed to expert comparisons adjusted their behaviour, submitting longer queries, clicking more deliberately, and using advanced techniques. %Social comparison offered concrete goals, helping users calibrate and refine their strategies.

Recent studies have also explored \textit{boost} interventions (see Section~\ref{sec:boosting}) \cite{hertwig2017nudging} to promote better search practices, with mixed outcomes. Some interventions successfully influenced behaviour, for example, reducing cookie acceptance \cite{ortloff2021effect} or fostering more balanced exploration of controversial topics \cite{bink2024balancing}. Others, such as 
attempts to increase intellectual humility during search \cite{rieger2024potential} or guide query reformulation through reflective prompts \cite{elsweiler2025query}, had weaker effects. Moraveji et al.~\cite{moraveji2011measuring} showed that in-task search tips could improve knowledge of search features, with effects persisting after one week.

Together, these works demonstrate that even without formal instructional goals, feedback mechanisms (via metrics, comparisons, or prompts) can help users develop more effective search strategies.

\subsection{Search as Learning}
\label{sec:sal}

The \textit{Search as Learning} (SAL) paradigm views search not just as a retrieval task, but as a cognitively active process through which users acquire new knowledge or skills. Câmera et al. \cite{camara2021searching} show that adding instructional scaffolding to web search changes search behaviour but does not improve learning gains, and that real-time feedback can even distract users from learning. Other work has focused on how behaviour changes as users gain domain knowledge, such as during courses or projects \cite{wildemuth2004effects,vakkari2003changes}. For example, students gradually adopt more specific vocabulary and improve their query reformulation strategies over time. Inspired by learning theory, researchers have analysed search tasks using Bloom’s taxonomy and other educational frameworks to understand different learning levels and their system requirements \cite{marchionini2006exploratory,jansen2009using,rieh2016towards}.

Learning theory has also been used to evaluate learning outcomes. Wilson and Wilson \cite{wilson2013comparison} applied a revised Bloom’s taxonomy to exploratory search. Bailey et al.~\cite{bailey2012user} used search logs to classify tasks, identifying categories like topic exploration and procedural learning. Eickhoff et al.~\cite{eickhoff2014lessons} extended this via automated analysis, finding that learning-related sessions accounted for a disproportionately large share of search time. Frummet et al.~\cite{frummet2024cooking} examined learning in agent-based search, measuring knowledge gain via the number of factual assertions (knowledge chunks) in responses.

While SAL has largely focused on \textit{searching to learn}, Rieh et al.~\cite{rieh2016towards} also draw attention to \textit{learning to search}. In addition to reviewing the former, they propose a research agenda for the latter, calling for information literacy programmes embedded in search system design. Their ideas include integrating scaffolding and guided inquiry to foster critical thinking and active engagement. 

We build on these ideas concretely by formally applying didactic frameworks to existing and emerging information access systems. We show how these systems already reflect, or could be redesigned to support, instructional principles such as scaffolding, feedback, and self-regulated learning.

\section{Didactic Frameworks}\label{sec:didactic}

%To explore how information access systems can function as instructional environments, we draw on a set of complementary frameworks from education, psychology, and behavioural science. These were selected for their relevance to user learning and behaviour within interactive systems, their use in prior HCI and IR work, and their conceptual diversity. While some (e.g., scaffolding, cognitive apprenticeship) are rooted in educational theory, others (e.g., nudging, boosting) originate in decision science but offer valuable insights into how systems can shape behaviour and support skill development.
We draw on seven complementary frameworks from education, psychology, and behavioural science that explain how systems can guide, support, or empower users during search (see Table \ref{tab:didactic-comparison}). Together, these frameworks span a continuum—from implicit, system-driven interventions to explicit, learner-driven models, offering a broad lens for analysing and designing information access systems with instructional goals in mind. Each subsection below outlines the framework’s origin, instructional model, pedagogical strengths and limitations, and its implications for system design.

\subsection{Nudging}

\textbf{Didactic model:} Implicit behavioural guidance through interface design.

\textbf{Key idea and origin:} Nudging stems from behavioural economics, especially the work of Thaler and Sunstein \cite{thaler2008nudge}. It describes interventions that steer user decisions without restricting choice. In information systems, nudging can take the form of result positioning \cite{zimmerman2019privacy}, or visual cues that encourage certain selections \cite{schwarz2011augmenting,yamamoto2011enhancing,elsweiler2017exploiting}.

\textbf{Pedagogical strengths:} Nudges require minimal user effort and can subtly guide users toward better behaviours, such as choosing more credible sources or reformulating vague queries. They work well under time pressure or cognitive load.

\textbf{Pedagogical limitations:} Nudges typically operate without explanation or reflection, which limits their value for building transferable skills or understanding. They may also be manipulative or ethically questionable if not transparent.

\textit{In didactic terms:} Nudging is behaviourist — it manipulates stimulus-response links without fostering conceptual understanding.

\subsection{Boosting}
\label{sec:boosting}

\textbf{Didactic model:} Explicit support for user competence through cognitive tools and strategies.

\textbf{Key idea and origin:} Boosting originates in decision science and psychology, particularly through the work of Hertwig and Grüne-Yanoff \cite{hertwig2017nudging}. Unlike nudging, boosting aims to strengthen users’ own abilities,  for example, by teaching how to evaluate information credibility, formulate effective queries, or compare sources. A related idea is self-nudging \cite{torma2018nudge}, in which users design their own choice environments to sustain desirable behaviours. This is closely aligned with the autonomy goals of boosting.

\textbf{Pedagogical strengths:} Boosting promotes autonomy, skill transfer, and long-term competence. It supports critical thinking and aligns well with educational goals of empowerment and self-directed learning.

\textbf{Pedagogical limitations:} Boosting often requires more effort and attention from users. Its effectiveness depends on motivation, prior knowledge, and the user's ability to apply the skill in context.

\textit{In didactic terms:} Boosting is constructivist: it builds mental models and strategies that learners can apply across contexts.

\subsection{Scaffolding and the Zone of Proximal Development (ZPD)}

\textbf{Didactic model:} Adaptive, temporary support within the learner’s developmental range.

\textbf{Key idea and origin:} The concept of the Zone of Proximal Development was introduced by Vygotsky \cite{vygotsky1978mind}, who argued that learning occurs most effectively in the space between what a learner can do independently and what they can do with support. Scaffolding, later formalised by Wood, Bruner, and Ross \cite{wood1976role}, refers to the instructional strategies that provide this support and fade over time as learners gain competence.

\textbf{Pedagogical strengths:} Scaffolding tailors support to the learner's current state, offering timely and context-sensitive assistance. It facilitates learning by doing and supports skill development in authentic tasks.

\textbf{Pedagogical limitations:} Effective scaffolding requires accurate detection of learner needs and progress. In systems design, this can be difficult to model and scale. Poorly timed or excessive scaffolding may hinder autonomy.

\textit{In didactic terms:} Scaffolding is socio-constructivist — it frames learning as a dynamic, supported process grounded in social and contextual interaction.

\subsection{Cognitive Apprenticeship}

\textbf{Didactic model:} Learning through guided experience, modelling, and gradual transfer of responsibility.

\textbf{Key idea and origin:} Cognitive apprenticeship was developed by Collins, Brown, and Newman \cite{collins1989cognitive} as an instructional approach that makes expert thinking visible. Learners acquire complex skills through modelling, coaching, articulation, reflection, and fading. This model is especially relevant for skills like search, evaluation, and critical reasoning. There are similarities between scaffolding and cognitive apprenticeship. Both support learning through guidance. However cognitive apprenticeship extends this idea by explicitly modelling expert thinking and gradually transferring strategic control to the learner.

\textbf{Pedagogical strengths:} Supports deep, situated learning and strategy transfer. Encourages learners to internalise expert practices and apply them independently. Aligns well with real-world, complex problem-solving.

\textbf{Pedagogical limitations:} Requires sustained interaction, modelling, and reflection — making it harder to implement in lightweight, time-sensitive systems. Can be resource-intensive and difficult to scale.

\textit{In didactic terms:} Cognitive apprenticeship is situated and developmental — it supports the gradual acquisition of complex practices through guided participation in authentic tasks.
\begin{table*}[t]
\centering
\caption{Comparison of Didactic Frameworks for Information Access System Design}
\begin{tabular}{p{2.8cm} p{3.6cm} p{3.2cm} p{4.2cm} p{2.8cm}}
\toprule
\textbf{Framework} & \textbf{Instructional Style} & \textbf{Role of the Learner} & \textbf{System Design Implications} & \textbf{Pedagogical Orientation} \\
\midrule
\textbf{Nudging} & Implicit behaviour guidance via environment structure & Passive decision-maker guided by defaults and cues & Subtle interface adjustments (e.g., result ordering, defaults) to steer user behaviour & behaviourist \\

\textbf{Boosting} & Explicit skill-building through cognitive tools & Active learner who applies strategies across tasks & Provide heuristics, explanations, or interactive prompts to foster competence & Constructivist \\

\textbf{Scaffolding (ZPD)} & Temporary, adaptive support within learner’s current limits & Guided learner who builds competence through assisted performance & Detect user ability and provide just-in-time, fading support & Socio-constructivist \\

\textbf{Cognitive Apprenticeship} & Learning through modelling, coaching, and fading & Participant in authentic tasks, learning expert-like strategies & Expose expert reasoning via LLM explanations or agent behaviour & Situated cognition \\

\textbf{Self-Regulated Learning (SRL)} & Support for planning, monitoring, and reflection & Self-monitoring learner who sets goals and adapts strategies & Provide feedback, history, and reflection tools for metacognitive control & Metacognitive constructivist \\

\textbf{Instructional Feedback} & Formative guidance for task and process improvement & Reflective learner who uses feedback to adjust behaviour & Use micro-feedback loops to reinforce effective strategies without overloading the user & Formative / cognitive \\

\textbf{Heutagogy} & Learner-directed, flexible, capability-focused instruction & Autonomous learner who determines their own path and level of support & Let users control depth/timing of guidance; offer opt-in learning tools or modes & Humanistic / andragogical \\
\bottomrule
\end{tabular}
\label{tab:didactic-comparison}
\end{table*}
\subsection{Self-Regulated Learning (SRL)}

\textbf{Didactic model:} Support for metacognitive control and self-directed strategy use.

\textbf{Key idea and origin:} Self-regulated learning emphasises learners’ ability to plan, monitor, and evaluate their learning process. Central to this model is the work of Zimmerman \cite{zimmerman2002becoming} and Pintrich \cite{pintrich2000role}, who conceptualise learning as an active, cyclical process involving goal-setting, strategic action, and self-reflection.

\textbf{Pedagogical strengths:} SRL fosters learner independence, promotes metacognition, and supports transfer of strategies across contexts. Systems based on this model can help users reflect on their behaviour and adjust future actions, e.g., through search trails, progress indicators, or prompts for reflection.

\textbf{Pedagogical limitations:} Requires motivation and metacognitive awareness from the user, which may not always be present. Without guidance, novice users may struggle to regulate their behaviour effectively, particularly under time pressure or in unfamiliar domains.

\textit{In didactic terms:} Self-regulated learning is metacognitive and cyclical — it supports learners in managing their own learning process through planning, monitoring, and reflection.

\subsection{Instructional Feedback Models}

\textbf{Didactic model:} Targeted, actionable, and timely feedback for performance and understanding.

\textbf{Key idea and origin:} Hattie and Timperley’s feedback model \cite{hattie2007power} defines effective feedback as information that helps learners answer three questions: “Where am I going?”, “How am I going?”, and “Where to next?” The model distinguishes between feedback that addresses tasks, processes, self-regulation, and the self, emphasising the importance of timing and specificity.

\textbf{Pedagogical strengths:} High-quality feedback supports both immediate correction and long-term understanding. In information access systems, feedback can take the form of micro-messages (e.g., “You considered multiple sources. Great strategy!”) or adaptive suggestions for alternative actions.

\textbf{Pedagogical limitations:} Feedback can be overwhelming, demotivating, or ignored if poorly timed or framed. It is also challenging to design feedback that balances usefulness with minimal cognitive disruption, especially in low-attention tasks.

\textit{In didactic terms:} Instructional feedback is formative and reflective — it aims to guide the learner’s current performance while fostering deeper understanding and future improvement.

\subsection{Heutagogy (Self-Determined Learning)}

\textbf{Didactic model:} Learner agency and flexible, capability-focused learning.

\textbf{Key idea and origin:} Heutagogy was introduced by Hase and Kenyon \cite{hase2000from}. It emphasises the development of learners' capacity to determine not only what and how to learn, but also how to evaluate their own learning needs. It values capability over mere competency, stressing adaptability and self-direction.

\textbf{Pedagogical strengths:} Heutagogy aligns well with diverse learners, lifelong learning, and personalised education. In information access systems, it supports giving users control over the timing and depth of instructional support (e.g., allowing users to toggle guidance modes or select the level of scaffolding they prefer).

\textbf{Pedagogical limitations:} Places high demands on learner autonomy, which may disadvantage those with limited prior knowledge or confidence. In system design, it may require complex personalisation infrastructure to adapt meaningfully to different user preferences.

\textit{In didactic terms:} Heutagogy is learner-centred and autonomy-driven. It positions users as self-determined agents in control of their own learning pathways.

Together, these seven frameworks represent different types and levels of learning support, each with varying demands in terms of user cognitive load and the amount of personal information required for effective personalisation.

\section{Case Studies}

To concretely demonstrate the value of applying didactic frameworks to the design and evaluation of information access systems, we have selected five case studies that represent a range of system types and instructional potentials. These include traditional search features, conversational generative AI, and agent-based systems. The selection was guided by the central goal of this paper: to explore how information access systems can act as \emph{instructional partners}, supporting not only task completion but also the development of users' information literacy through interaction. The examples reflect diversity in interaction modality and didactic strength, and to include both strong teaching systems and a useful contrast case.

\begin{figure*}[t]
  \centering
  \includegraphics[width=\textwidth]{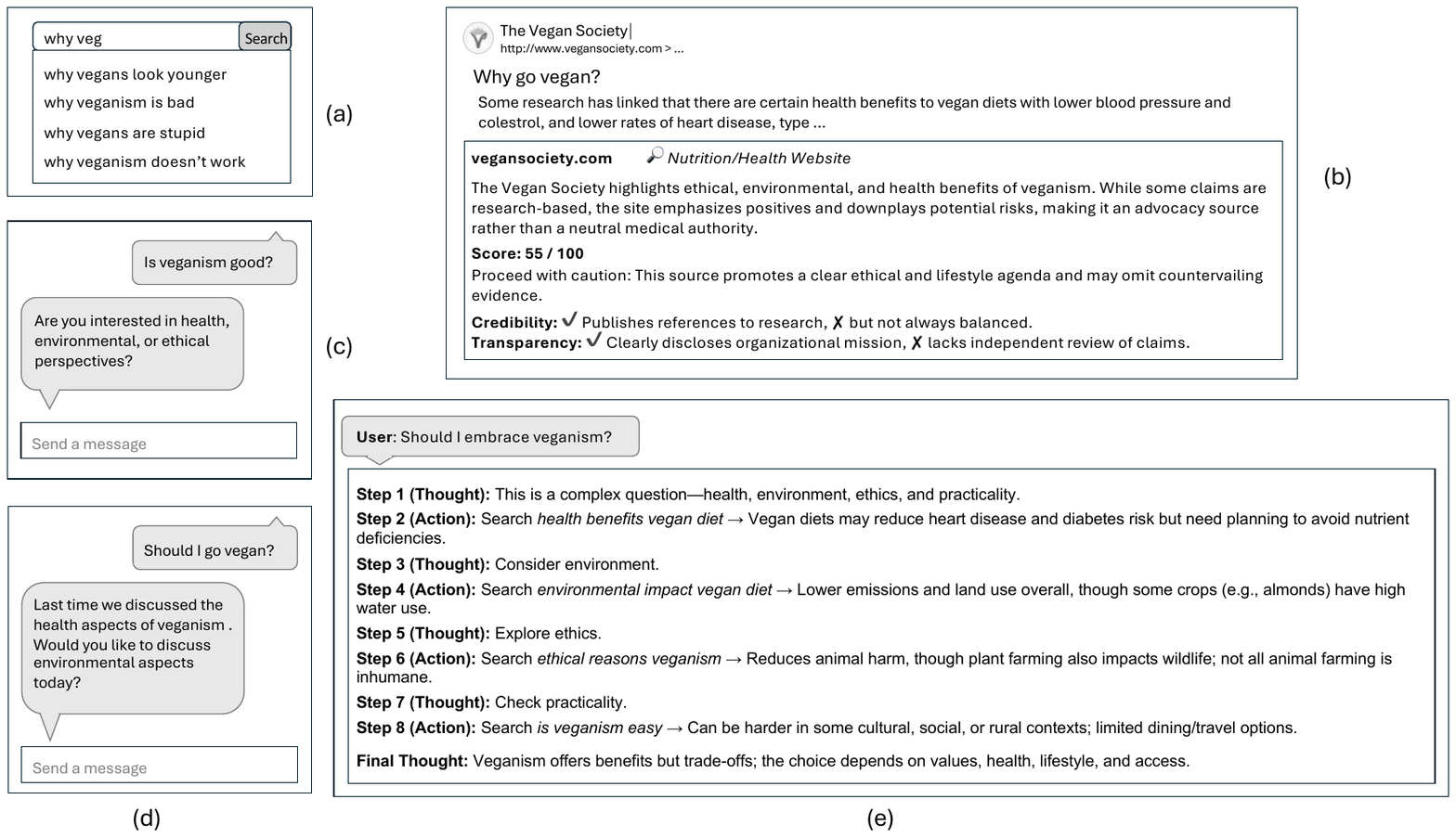}
  \caption{Case study mockups: (a) query suggestion, (b) source labels, (c) GenAI-Chat with clarifying questions, (d) GenAI-Chat with memory, (e) ReAct Planning Agent}
  \label{fig:mockups}
\end{figure*}

\begin{itemize}
    \item \textbf{Query Auto-Completion} (Figure~\ref{fig:mockups}(a)) is a common feature in modern search engines that offers query suggestions as users type, typically based on popularity, trends, or contextual cues \cite{cao2008context}. It can reduce typing effort and guide users, especially novices \cite{niu2014use}, towards query formulations. However, it provides no explanation or feedback, and the interaction is brief and one-way, offering little opportunity for deeper learning or strategy development.

    \item \textbf{Source Labels and Fact-Check Annotations}, such as “Highly Cited,” “Fact Checked by...,” or “From a known medical source”, provide visual indicators intended to inform judgments of credibility and trustworthiness \cite{dennis2023search}(Figure \ref{fig:mockups}(b)). These markers are often displayed alongside individual search results. Users are not required to engage with them explicitly, and the system does not always explain their meaning or origin in context. Their effectiveness depends on users noticing, understanding, and acting upon them.

    \item \textbf{GenAI-Chat with Clarifying Questions} refers to LLM-based conversational systems that, when faced with vague or underspecified queries, prompt the user to clarify their intent (Figure \ref{fig:mockups}(c)). For example, when asked “Is veganism good?”, the system might follow up with “Do you mean for the environment, for health, or for animals?” These clarifying turns can help users refine their goals and produce more focused queries. The interaction is dialogic and lightweight, with the system adapting its prompts based on initial input.

    \item \textbf{GenAI-Chat with Memory and Session Reflection} represents a more persistent, context-aware form of interaction. In systems with memory, prior queries and conversations can influence current responses (Figure \ref{fig:mockups}(d)). Some implementations allow users to view, rename, or delete previous sessions, and may even present summaries of past discussions. This continuity allows users to reflect on earlier explorations, maintain coherence across sessions, and gradually refine their understanding of a topic over time.

    \item \textbf{ReAct Planning Agents} are agentic systems that interleave reasoning and acting, typically by combining language model reasoning with external tool use (e.g., web search)\cite{yao2023react} (Figure \ref{fig:mockups}(e)). Such agents decompose tasks into substeps, perform actions such as retrieving information or calculating results, and iteratively refine their approach. Although the system proposed in \cite{yao2023react} was designed to operate autonomously, its intermediate thoughts and actions are surfaced to users. When revealed, these traces show how complex queries are solved, but users cannot control the agent’s plan or reasoning, limiting engagement and reflection without additional scaffolding.
\end{itemize}

Together, these five examples provide a balanced and multi-faceted illustration of how systems can, implicitly or explicitly, support users not only in finding information but in \emph{becoming better information seekers}.

\section{Illustrative Search Tasks}
\label{sec:tasks}

To illustrate how different information access systems embody distinct instructional roles, we focus on two representative search tasks that differ in complexity, cognitive demand, and relevance to everyday users. These tasks were selected not only because they are commonly studied in IIR but since expose the ways in which systems may (or may not) support users in refining their strategies, developing their information literacy, and achieving a more accurate or thoughtful understanding of the topic at hand.

\subsection{Task A: Validating a Common Claim}

\textit{“Is it true that drinking coffee stunts your growth?”} This task reflects a common type of information-seeking behaviour: users encounter a familiar but questionable claim and wish to verify its truth. Such tasks are studied in relation to health issues \cite{white2013beliefs,pogacar2017positive,bink2022featured,mayerhofer2025blending}. While the query appears simple, successful resolution requires users to formulate precise search terms, assess the credibility of sources, and reconcile conflicting information, making the task pedagogically rich despite its surface simplicity.

This task is especially well-suited to systems that can scaffold critical questioning, model effective verification strategies, or promote source evaluation. It allows us to observe whether systems merely present plausible content or actively guide the user toward reliable evidence.

\subsection{Task B: Taking a Stance on a Debated Topic}

\textit{“Should I embrace veganism?”} This task reflects a common type of complex information-seeking behaviour in which users explore a controversial or value-laden topic with the intention of forming or revising their stance (e.g. \cite{draws2023viewpoint,rieger2024responsible}. It involves the evaluation of scientific evidence, ethical arguments, environmental implications, and personal health considerations. Search results are likely to vary widely in framing, tone, and credibility, depending on the user’s initial phrasing and search path.

Such tasks are pedagogically rich because they require more than factual recall. They demand critical evaluation, perspective-taking, and synthesis of diverse viewpoints \cite{rieger2024responsible}. These searches can meaningfully impact users' opinions, self-concept, and subsequent behaviours.

From an instructional standpoint, this task enables the analysis of how systems support or hinder open-minded exploration. Effective systems might scaffold multiperspectivity, prompt the comparison of pros and cons, or model how to weigh competing types of evidence. Conversely, systems that personalise results too quickly or fail to challenge initial assumptions risk reinforcing echo chambers or preexisting biases.

This task allows us to assess whether systems help users develop not just conclusions, but also the competencies needed for informed and reflective decision-making.
\section{Instructional Walkthroughs }
\label{sec:walkthrough}

To illustrate how each system or feature supports (or neglects) instructional opportunities, we provide walkthroughs of user interactions across two tasks: validating a common claim and forming a stance on a debated issue. These narratives highlight how different interaction patterns align with or fall short of didactic principles.

\subsection{Query Auto-Completion (Search)}

Query auto-completion influences how users frame both simple and complex questions. 
When verifying whether coffee stunts growth, suggestions such as ``does coffee stunt growth'' or ``does coffee help you lose weight'' reflect popular myths rather than conceptual refinement. 
Similarly, when exploring whether to embrace veganism, completions such as ``why veganism is bad'' or ``why vegans are wrong'' may introduce biased framings drawn from aggregate user behaviour. 
While these cues shape query formulation, they offer no guidance or feedback for reflection, limiting opportunities for users to develop better search strategies or awareness of bias.

\subsection{Source Labels and Fact-Check Annotations}

Source labels such as ``Fact Checked,'' ``Highly Cited,'' or ``From a known medical source'' provide quick credibility cues across both simple and complex searches. 
In the coffee example, such tags may guide users toward trustworthy results, and in the veganism task they can help distinguish factual reports from opinion pieces. 
However, because these labels rarely explain their origin or meaning, users must interpret their significance independently, and the lack of feedback limits reflection and deeper learning.

\subsection{GenAI-Chat with Clarifying Questions}

Conversational systems that pose clarifying questions can scaffold user understanding across tasks. 
When asked ``Does coffee stunt your growth?'' the system may prompt for context, such as age group, and briefly explain why clarification matters. 
In more open-ended questions such as ``Should I go vegan?'', it may ask whether the user is interested in health, environmental, or ethical aspects. 
By modelling expert inquiry and highlighting dimensions of a topic, such systems encourage more precise formulation and support metacognitive reflection.

\subsection{GenAI-Chat with Memory and Reflection}

Systems that maintain conversational memory allow users to connect ideas across sessions. 
When verifying information, references to earlier queries about related topics (e.g., caffeine or nutrition myths) promote continuity and reinforce reasoning. 
For complex stance formation, such as evaluating veganism, the system may summarise previous discussions or suggest revisiting arguments already considered. 
These features support self-regulated learning by enabling users to track their evolving understanding and reasoning process over time.

\subsection{ReAct Planning Agent}

Agentic systems such as ReAct decompose tasks into sub-steps and expose their reasoning process. 
When validating claims like the coffee example, the agent defines the statement, retrieves and compares evidence, and displays its intermediate steps. 
For more complex decisions such as adopting veganism, it may plan sub-questions (e.g., health, ethics, environment) and present contrasting viewpoints. 
By making reasoning visible and structured, the system models reflective problem-solving and helps users learn transferable verification strategies.

The walkthroughs above illustrate how different systems and features shape user behaviour across tasks. 
While some afford structured guidance or reflective cues, others exert more subtle influence through interface design. 
Yet to fully understand their instructional value, we must go beyond describing behaviour and examine these systems through an explicitly didactic lens. 
In the next section, we apply seven instructional frameworks to interpret how these technologies support the development of information literacy.

\section{Didactic Framework Analysis}
\label{sec:analysis-didactic}

While many information access systems are designed to support users in completing tasks efficiently, far fewer are evaluated in terms of their capacity to foster learning, reflection, or skill development. To examine the instructional potential of different systems, we analyse five representative features and AI-based interfaces through the lens of established didactic frameworks. These frameworks provide distinct perspectives on how systems might guide, support, or limit user development during search and inquiry. Each subsection below highlights one framework, explains its pedagogical foundation, and evaluates which systems align with its principles based on the walkthroughs provided above. This analysis enables a deeper understanding of how information access systems function not only as tools, but as \textit{teachers}.

\begin{table*}[t]
\centering
\caption{Instructional alignment of systems and features across didactic frameworks.}
\label{tab:framework-alignment}
\begin{tabular}{lccccccc}
\toprule
\textbf{System / Feature} & \textbf{Nudging} & \textbf{Boosting} & \textbf{Scaffolding} & \textbf{Cog. Apprent.} & \textbf{SRL} & \textbf{Instr. Feedback} & \textbf{Heutagogy} \\
\midrule
Query Auto-Completion (Search) & \checkmark & -- & -- & -- & -- & -- & -- \\
Source Labels / Fact-Check Tags & \checkmark & $\pm$ & -- & -- & $\pm$ & -- & $\pm$ \\
GenAI-Chat with Clarifying Questions & $\pm$ & $\pm$ & \checkmark & \checkmark & $\pm$ & $\pm$ & -- \\
GenAI-Chat with Memory + Reflection & -- & $\pm$ & $\pm$ & $\pm$ & \checkmark & \checkmark & \checkmark \\
ReAct Planning Agent & -- & $\pm$ & $\pm$ & \checkmark & \checkmark & \checkmark & $\pm$ \\
\bottomrule
\end{tabular}

\vspace{0.5em}
\raggedright
\textit{Legend:} \checkmark = Strong support, $\pm$ = Partial or indirect support, -- = Not supported.
\end{table*}

\subsection{Nudging}

Nudging operates through subtle changes in choice architecture that steer user behaviour without explicit instruction. Both \textbf{query auto-completion} and \textbf{source labels} clearly function as nudges, shaping user input and trust decisions through interface design. For example, in the veganism task, the autocomplete feature surfaces suggestions like ``why veganism is bad'' or ``why vegans are wrong,'' potentially reinforcing biased framings. Similarly, source labels such as ``Fact Checked'' or ``Highly Cited'' in the coffee claim task can shift user attention towards certain results without helping them understand why those sources are reliable. Nudging is fast and effective, but from a didactic standpoint, it is purely behaviourist: it shapes behaviour without fostering conceptual understanding or reflection.

\subsection{Boosting}

Boosting aims to improve users’ decision-making skills by giving them transparent cues or tools to make better judgments. \textbf{Source labels} may serve this role if users understand their origin and meaning, such as distinguishing peer-reviewed sources from opinion blogs when evaluating whether coffee stunts growth. \textbf{GenAI-Chat with clarifying questions} can also promote boosting by repeatedly prompting users to disambiguate their queries—for example, distinguishing between ethical, health, or environmental motivations when asking about veganism. However, because most systems do not reinforce or explain these cues, their value depends heavily on users’ prior knowledge and engagement. Boosting supports skill development, but requires interpretive effort from the learner to be pedagogically effective.

\subsection{Scaffolding (Zone of Proximal Development)}

Scaffolding supports learners in performing tasks just beyond their current competence through temporary, adaptive support. \textbf{GenAI-Chat with clarifying questions} exemplifies this by helping users refine vague queries. In the coffee task, it prompts clarification such as ``Are you asking about children or adults?'', enabling more targeted information seeking. In the veganism stance task, it may ask ``Are you concerned about health, ethics, or the environment?'', thereby helping users structure their inquiry. \textbf{ReAct planning agents} scaffold complex decision-making by decomposing the veganism task into subtasks such as evaluating scientific findings or ethical arguments. However, this support does not adapt to user progress or fade over time. As such, ReAct agents scaffold the task rather than the learner. In true didactic terms, scaffolding requires sensitivity to the learner's evolving competence and a gradual reduction of support.

\subsection{Cognitive Apprenticeship}

Cognitive apprenticeship involves making expert thinking visible—teaching not just answers, but how to think through problems. \textbf{ReAct planning agents} embody this by exposing reasoning steps when validating the coffee claim: they may retrieve definitions, compare studies, and explicitly walk through logical steps. In the veganism task, they might contrast ethical, environmental, and nutritional dimensions, showing how to weigh evidence and reach conclusions. \textbf{GenAI-Chat with clarifying questions} models expert inquiry by asking users to specify assumptions or contexts. These interactions give users insight into expert strategies that can be generalised to other information tasks. In contrast, static features such as \textbf{query suggestions} or \textbf{source labels} provide no visibility into the reasoning behind them. Cognitive apprenticeship supports transfer by modelling, not just delivering, expertise.

\subsection{Self-Regulated Learning}

Self-regulated learning (SRL) refers to a learner’s ability to monitor, evaluate, and adapt their strategies over time. \textbf{GenAI-Chat with memory and reflection} supports SRL by summarising prior interactions and suggesting continuity. For example, in the veganism stance task, the system might reference previous queries about animal welfare or climate impact, prompting the user to integrate these into their current reasoning. \textbf{ReAct agents} promote SRL by structuring subtasks and showing how planning and evidence gathering unfold, helping users learn how to regulate their own inquiry processes. \textbf{Source labels} may contribute marginally to SRL if users begin to reflect on credibility cues over time. Most conventional search systems, however, provide no support for tracking conceptual growth or revisiting earlier queries. SRL is essential for metacognitive development and transfer across contexts.

\subsection{Instructional Feedback}

Instructional feedback provides users with information on the quality of their decisions or inputs. \textbf{GenAI-Chat with memory and reflection} and \textbf{ReAct agents} offer explicit and contextualised feedback by revisiting prior steps or offering improved reasoning paths. For instance, after an initial vague veganism query, the system might summarise what the user explored, or suggest refining the query focus. In the coffee claim task, a ReAct agent might revise its own plan after encountering weak evidence, modeling how to course-correct. \textbf{Clarifying questions} also offer implicit feedback by signaling that the user's input lacks specificity. In contrast, features like \textbf{autocomplete} and \textbf{source labels} do not evaluate or respond to the user's actions in a way that supports learning. Instructional feedback is vital for iterative learning and developing effective strategies.

\subsection{Heutagogy}

Heutagogy emphasises self-directed, learner-defined inquiry, where users shape their own learning paths and goals. \textbf{GenAI-Chat with memory and reflection} aligns closely with this model, allowing users to revisit, revise, and direct their own trajectory over time. In the veganism task, the system may enable users to steer the conversation based on evolving priorities, ethical, social, or health-related, and choose which threads to pursue. \textbf{ReAct agents} can support heutagogy if designed to accept user-defined goals and defer to user priorities during task decomposition. Static features such as \textbf{query suggestions} and \textbf{source labels} offer no such user control. Heutagogy requires systems to respect user agency, support personalisation, and maintain persistent context over time.

\textbf{Synthesis:} Traditional features such as query suggestions and source labels influence user behaviour but provide little support for learning. In contrast, conversational and agentic systems align more strongly with scaffolding, cognitive apprenticeship, and self-regulated learning principles. However, even these more advanced systems rarely adapt to users' evolving competencies or explicitly fade support over time.

%Across frameworks, we observe that current information access systems vary widely in their instructional affordances. Although features such as query suggestions and source labels influence user behaviour, they offer limited support for learning or strategy development. In contrast, AI-based systems, particularly those that model reasoning steps, adapt across sessions, or prompt reflection, demonstrate stronger alignment with principles of scaffolding, cognitive apprenticeship, and self-regulated learning. However, even these more advanced systems rarely adapt to users' evolving competencies or explicitly fade support over time. This analysis suggests that treating information access systems as didactic tools reveals both promising directions and current gaps, especially in their ability to foster transferable search literacy and critical thinking.

\subsection*{Applying Frameworks to System Design}

To illustrate the design potential of these frameworks, we present four hypothetical information access systems, each intentionally designed around the core principles of a specific instructional model. These examples demonstrate how didactic concepts can guide not only the evaluation but also the design of future search and reasoning systems.

\textbf{Scaffolding (ZPD)}.  
Consider a system designed to support a user researching whether veganism is healthy. Upon entering a vague query like ``Is veganism healthy?'', the system prompts the user to refine their question, offering distinctions such as ``Are you interested in long-term health outcomes, nutrient balance, or weight loss?'' It then surfaces curated overviews, prompts the user to compare sources, and scaffolds more complex operations such as identifying expert consensus. This will be cognitively burdening for the user and may slow progress and frustrate. Over time, as the system detects improvements in the user's query formulation and evaluation behaviour, it reduces its guidance and returns greater autonomy. The support is gradually withdrawn in response to observed competence—realising the core principle of fading in scaffolding.

\textbf{Instructional Feedback}.  
A feedback-oriented system might be designed to help users reflect on their information evaluation decisions. When a user selects a dubious or one-sided article, the system provides a gentle prompt: ``This source has been flagged for bias. Would you like to see how other sources report on the same issue?'' It also offers just-in-time feedback on search strategies, e.g., ``You’ve only viewed sources from one domain—would you like to diversify?'' Such feedback supports formative learning without penalising users, helping them iteratively improve their search behaviour.

\textbf{Cognitive Apprenticeship}.  
A system built around cognitive apprenticeship could include an ``expert trace'' mode. As users search, they can optionally follow paths taken by domain experts responding to the same query—seeing what sub-questions they asked, which sources they evaluated, and how they made credibility judgments. Explanatory overlays show why certain steps were taken, making expert reasoning visible. Over time, users can compare their strategies to those of more experienced searchers and begin to internalise those patterns.

\textbf{Self-Regulated Learning}.  
An SRL-oriented system would support metacognitive reflection across sessions. For example, after a session on veganism and health, the system might summarise: ``Last time, you focused on ethical arguments. Would you like to continue exploring nutritional perspectives?'' Users could revisit their past search paths, annotate sources, and set goals for future sessions. The system supports monitoring and planning, enabling users to take ownership of their learning trajectory while maintaining continuity over time.
\section{Discussion and Limitations}
\label{sec:discussion}

Reframing information access systems as instructional environments opens new perspectives on system design and evaluation. 
Instead of focusing solely on immediate relevance or satisfaction, we can ask whether users improve as searchers, developing transferable strategies, critical reasoning skills, and greater epistemic awareness.

Our analysis suggests that current systems vary widely in their instructional affordances. 
Traditional features such as query auto-completion and source labels influence user behaviour through nudging but offer limited support for deeper learning or strategy development. 
In contrast, AI-based systems, especially those that model reasoning steps, enable reflection, or adapt across interactions, align more strongly with didactic frameworks such as scaffolding, cognitive apprenticeship, and self-regulated learning. 
The conversational nature of these systems is certainly advantageous and could be designed to act like a personalised librarian or coach.

However, even these advanced systems fall short of their instructional potential. 
They rarely diagnose user knowledge, personalise support based on learning progress, or fade assistance over time—core tenets of effective pedagogy. 
Their capacity to support user learning therefore remains largely implicit and underutilised. 
Rather than prescribing user behaviour, these systems can invite reflection and self-directed improvement, aligning with heutagogical and autonomy-respecting principles. 
Instructional design in information access should emphasise agency and metacognition over control, helping users become active participants in their own learning process.

To move forward, didactic frameworks should not only be used to evaluate systems but also to inspire their design. 
In Section~\ref{sec:analysis-didactic}, we presented several hypothetical systems intentionally built to embody the core principles of specific frameworks. 
These examples illustrate how search features, conversational interfaces, and AI agents could provide clarifying prompts, formative feedback, or reflective tools that adapt over time to user development.

Nonetheless, designing for learning involves inherent trade-offs. 
Systems that aim to teach users how to search better, through prompts, scaffolds, or reflection, may temporarily increase cognitive load and reduce efficiency. 
The appropriate level of instructional support will depend on task complexity, user goals, and prior knowledge. 
Personalised and context-aware adaptation will be crucial to balancing learning opportunities with usability and satisfaction.

A further tension lies in the assumption that users must learn to resist exploitative or attention-maximising design. 
If we can imagine systems that teach users to be more discerning, we should also imagine platforms that are less manipulative by design. 
These directions are not mutually exclusive: empowerment-oriented systems can coexist with structural reforms that promote ethical and transparent information infrastructures. 
Examples from other domains—such as navigation systems that teach route planning, grammar tools that model correct syntax, or energy dashboards that visualise consumption to promote awareness—illustrate how infrastructure itself can act as a teacher. 
Such analogies highlight the broader societal potential of pedagogically aware information systems to foster resilience, critical thinking, and informed agency in digital environments.

\subsection*{Limitations}

The presented analysis is necessarily selective. For space reasons, we focused on a small set of instructional frameworks and search system features, chosen to illustrate how pedagogical perspectives can be applied across both traditional interfaces and emerging generative systems. Other didactic models, such as constructivism, direct instruction, or situated learning, may have provided additional or contrasting insights. Similarly, our case studies emphasise illustrative coverage rather than exhaustive typology.

While generative AI and agentic systems are prominent in our examples, this should not be taken as an uncritical endorsement. We are mindful of known risks such as hallucinations, inconsistent reasoning, and opaque behaviour (see Shah and Bender \cite{shah2022situating}—issues that our chosen frameworks do not explicitly address. Importantly, we do not claim that only advanced systems can support instructional goals. Many of the characteristics highlighted by the frameworks, such as feedback, scaffolding, or reflective prompts, could be achieved through traditional search interfaces, perhaps via well-timed messages, contextual help, or integration with lightweight generative agents.

Our aim is not to advocate for specific technologies, but to demonstrate the value of treating information access systems as potential teachers, and to encourage future research that embraces this perspective.

\section{Conclusion and Future Work}
\label{sec:conclusions}

This paper reframes information access systems as instructional environments that can support user development, not just information retrieval. Drawing on didactic frameworks from education, psychology, and behavioural science, including scaffolding, cognitive apprenticeship, nudging, boosting, and self-regulated learning, we analysed how existing systems and features implicitly or explicitly guide users. We proposed a conceptual foundation for evaluating and designing systems based on their instructional role, illustrated through concrete case studies and user walkthroughs.

Viewing search and generative AI systems through this lens opens new research directions at the intersection of learning theory, HCI, and interactive IR. We conclude by highlighting three promising directions for future work:

\begin{itemize}
    \item \textbf{Designing for Visible Expertise.} Future systems can be designed to transparently model expert reasoning, enabling users to observe, internalise, and adopt effective search and evaluation strategies. This supports deeper learning through cognitive apprenticeship.

    \item \textbf{Developing Adaptive, Fading Scaffolds.} Research is needed on personalised scaffolding mechanisms that adapt to user needs and gradually reduce support over time, promoting independence and skill transfer without increasing cognitive load.

    \item \textbf{Evaluating Instructional Value.} Existing evaluation methods often focus on relevance and satisfaction. A future research agenda should include new metrics and study designs to assess instructional impact (e.g., improvements in query formulation, source evaluation, or information literacy over time).
\end{itemize}

By rethinking information access systems as teachers, not just tools, we can design technologies that empower users to become more strategic, reflective, and resilient in their interactions with information.

\bibliographystyle{ACM-Reference-Format}
\balance
\bibliography{bibliography}

\end{document}